\documentclass[prd,twocolumn,nopacs,floatfix,amsmath,amssymb,floatfix,nofootinbib]{revtex4}
\usepackage{graphicx,dcolumn,booktabs,bm}
\usepackage{longtable,lscape}
\usepackage{txfonts}
\usepackage{overpic}
\usepackage{amssymb}
\usepackage{indentfirst}
\usepackage{feynmf}   
\usepackage{slashed}  
\usepackage{cases}
\usepackage{color,ulem}
\usepackage[
colorlinks, 
pdfborder=001,   
citecolor=blue
]{hyperref}

\graphicspath{{Figures/}} 

\begin{document}

	
\title{Production of charmed baryon $\Lambda_c(2860)$ via low energy antiproton-proton interaction}

\author{Qing-Yong Lin$^{1}$\footnote{corresponding author}}\email{qylin@jmu.edu.cn}
\author{Xiang Liu$^{2,3,4}$}\email{xiangliu@lzu.edu.cn}

\affiliation{
$^1$Department of Physics, School of Science, Jimei University, Xiamen 361021, China\\
$^2$School of Physical Science and Technology, Lanzhou University, Lanzhou 730000, China\\
$^3$Lanzhou Center for Theoretical Physics, Key Laboratory of Theoretical Physics of Gansu Province and Frontiers Science Center for Rare Isotopes, Lanzhou University, Lanzhou 730000, China\\
$^4$Research Center for Hadron and CSR Physics, Lanzhou University \& Institute of Modern Physics of CAS, Lanzhou 730000, China
}
	
\date{\today}
	
\begin{abstract}
	In the present work, we study the production problem of the charmed baryon $\Lambda_c(2860)^+$ at $\overline{\mbox{P}}$ANDA. With the $J^P=\frac{3}{2}^+$ assignment to $\Lambda_c(2860)^+$, an effective Lagrangian approach is adopted to calculate the cross section of $p\bar{p} \to \Lambda_c^-\Lambda_c(2860)^+$. The Dalitz plot analysis and the $D^0p$ invariant mass spectrum distribution are also given for the $p\bar{p} \to \Lambda_c^-\Lambda_c(2860)^+ \to \Lambda_c^-pD^0$ process. The numerical results show that the total cross section may reach up to about 10 $ \mu $b. With the designed luminosity of $\overline{\mbox{P}}$ANDA ($2\times10^{32}~\mbox{cm}^{-2}\mbox{s}^{-1}$), about $ 10^8 $ $\Lambda_c(2860)$ events can be expected per day by reconstructing the final $ pD^0 $.
\end{abstract}

\maketitle
	
\section{Introduction}\label{sec:intro}

Charmed baryon family has special place in whole hadron spectroscopy. Focusing on the charmed baryon, theorist and experimentalist have 
paid a lot of effort to reveal their nature, which has close relation to construct charmed baryon family and find out 
exotic state (see review articles \cite{Klempt:2009pi,Chen:2016spr,Guo:2017jvc,Kato:2018ijx,Galkin:2020iat} for the progress). 
In the past few years, charmed baryon family has become more and more abundant with the observation of more excited states of charmed baryon, which can be reflected by the present status of charmed baryons listed in Particle Data Group (PDG) \cite{Zyla:2020zbs}. 

Among these observed charmed baryons, $\Lambda_c(2860)^+$ was reported by LHCb in the $D^0$p channel \cite{Aaij:2017vbw}. The measured mass and width are
\begin{eqnarray*}
	M&=&2856.1^{+2.0}_{-1.7}\,(stat)\pm0.5(syst)^{+1.1}_{-5.6}(model)\, \mathrm{MeV},\\
	\Gamma&=&67.6^{+10.1}_{-8.1}(stat)\pm1.4(syst)^{+5.9}_{-20.0}(model)\, \mathrm{MeV},
\end{eqnarray*}
respectively. It should be noted that, the uncertainty related to the amplitude model was considered in the analysis presented in the LHCb paper \cite{Aaij:2017vbw}. Thus, when giving the experimental data of mass and width of $ \Lambda_c(2860)^+ $, this uncertainty was listed. This model uncertainty is marked by “(model)” in the resonant parameter. Meanwhile, the available experimental analysis indicates that the observed $\Lambda_c(2860)^+$ state has the spin-parity $J^P=\frac{3}{2}^+$.
Before the discovery of $\Lambda_c(2860)^+$, many studies were carried out on the mass spectra of charmed baryons with different scenarios and a $D$-wave charmed baryon with a mass around 2.85 GeV was predicted \cite{Chen:2014nyo,Chen:2016iyi,Lu:2016ctt,Chen:2016phw}, which is consistent with the observation of $\Lambda_c(2860)^+$ from LHCb \cite{Aaij:2017vbw}. Since then, the newly observed $\Lambda_c(2860)^+$ attracted more and more attentions from theorists to decode its inner structure and decay property. Lanzhou group furthermore studied the mass spectrum of the $\lambda$-mode excited charmed and charmed-strange baryon states \cite{Chen:2017aqm}. Their results indicate that $\Lambda_c(2860)^+$ associated with former $\Lambda_c(2880)^+$ can form a $D$-wave doublet [$3/2^+$,$5/2^+$]. More discussions can be found in Refs. \cite{Wang:2017vtv,Mao:2018eti,Faustov:2018vgl,Gandhi:2019xfw,Chen:2021eyk}, where different phenomenological methods/models were applied.

Besides the investigation of the mass spectrum, the strong decay properties of the low-lying D-wave charmed baryons were also studied within some methods. The $ D^0 p $ decay channel is of great importance to provide information on the inner structure of these highly excited $ \Lambda_c $. The authors carried out the two-body Okubo-Zweig-Iizuka-allowed decays of $\Lambda_c(2860)^+$ within the $^3P_0$ model and gave the decay widths of different processes \cite{Chen:2017aqm}. 
It is obvious that the $DN$ branching fraction reaches up to 95.6\%. A smaller value of 75\% was obtained with a similar model in Ref. \cite{Gong:2021jkb}. These results indicate that the $DN$ channel should be the major decay channel. However, the authors pointed that due to the limited phase space, the partial widths decaying into the $DN$ channel should be small \cite{Lu:2019rtg}.

In addition, the $ \Sigma_c\pi $ decay channel should be a channel of concern. However, no signal has been measured for $ \Lambda_c(2860) $ experimentally to date. We notice that the partial decay width and the ratio $\mathcal{R}=\frac{\Gamma[\Sigma_c(2520)\pi]}{\Gamma[\Sigma_c(2455)\pi]}$ have been predicted theoretically. 
The ratio $\mathcal{R}=0.47$ was predicted in Ref. \cite{Chen:2017aqm}. Similarly with the $^3P_0$ model, the authors obtained a value range from 2.8 to 3.0 \cite{Guo:2019ytq}. By virtue of a constituent quark model, Yao {\it et al.} got the partial widths to be 4.57 MeV and 0.95 MeV for the decay modes $|^2D_{\lambda\lambda}\frac{3}{2}^+\rangle \to \Sigma_c\pi$ and $|^2D_{\lambda\lambda}\frac{3}{2}^+\rangle \to \Sigma_c^*\pi$, respectively \cite{Yao:2018jmc}, which leads to a value $\mathcal{R}=0.21$. Considering the measured width, the relatively large branching fractions indicate that $\Lambda_c(2860)^+$ might be observed in the $\Sigma_c(2455)\pi$ and $\Sigma_c(2520)\pi$ channels as well. In addition, it was pointed out that the ratio $\mathcal{R}$ for the nearby state $\Lambda_c(2880)$ may be strongly affected by $\Lambda_c(2860)$ \cite{Yao:2018jmc}. Thus, for the purpose of deeply understand $\Lambda_c(2860)$ and other charmed baryons, it is very important to measure the branching ratio. However, the current available information of $\Lambda_c(2860)^+$ is the measured mass and width from the channel decaying to $D^0p$. Consequently, more experimental information are strongly required to further understand the decay behavior of $\Lambda_c(2860)^+$. The ratio $\mathcal{R}=\frac{\Gamma[\Sigma_c(2520)\pi]}{\Gamma[\Sigma_c(2455)\pi]}$ for the nearby state $\Lambda_c(2880)$ may be strongly affected by $\Lambda_c(2860)$ \cite{Yao:2018jmc}.

Just reviewed above, the study of $\Lambda_c(2860)^+$ mainly emphasize its mass and decay \cite{Chen:2014nyo,Chen:2016iyi,Lu:2016ctt,Chen:2016phw,Chen:2017aqm,Wang:2017vtv,Mao:2018eti,Faustov:2018vgl,Gandhi:2019xfw,Chen:2021eyk,Gong:2021jkb,Lu:2019rtg,Guo:2019ytq,Yao:2018jmc}. The discussion of the production of 
$\Lambda_c(2860)^+$ is still absent. 
Until now, $\Lambda_c(2860)^+$ was only observed in the Cabibbo-favoured decay $\Lambda_b \to D^0p\pi^-$, where another two states $\Lambda_c(2880)^+$ and $\Lambda_c(2940)^+$ previously observed by the BaBar experiment \cite{Aubert:2006sp} were also confirmed \cite{Aaij:2017vbw}. Thus, it is interesting in exploring the $\Lambda_c(2860)^+$ production in other processes. As indicted by the strong decay behavior of $\Lambda_c(2860)^+$ \cite{Chen:2017aqm,Gong:2021jkb}, the $DN$ decay channel is dominant, which inspires our interesting in exploring the $\Lambda_c(2860)^+$ production via the low energy antiproton-proton interaction. We notice that
the future facility $\overline{\mbox{P}}$ANDA will exploit the annihilation of antiprotons with protons and nuclei to study the fundamental forces in nature \cite{Lutz:2009ff}. Study of the charmed baryon is one of the main physics goals of $\overline{\mbox{P}}$ANDA. Some parallel theoretical investigations were previously implemented on the productions of charmed baryons in antiproton-proton collisions \cite{He:2011jp,Lin:2014jza,Shyam:2017gqp,Haidenbauer:2016pva}. The above reason also pushes us to study the discovery potential of $\Lambda_c(2860)^+$ at $\overline{\mbox{P}}$ANDA, which can provide valuable information to future experimental exploration of $\Lambda_c(2860)^+$ at $\overline{\mbox{P}}$ANDA.

This work is organized as follows. After the Introduction, we present the theoretical model and the corresponding calculation details in Sec. \ref{sec2}. The numerical results of the $\Lambda_c(2860)^+$ production at $\overline{\mbox{P}}$ANDA will be given in Sec. \ref{sec3}, including the cross sections, the Dalitz plot and the $pD^0$ invariant mass spectrum. Finally, this paper ends with a discussion and conclusion (see Sec. \ref{sec4}).

\section{$\Lambda_c(2860)$ production in $\bar{p}p$ annihilation}\label{sec2}

\begin{figure}[htb]
	\begin{center}
		\scalebox{0.9}{\includegraphics[width=\columnwidth]{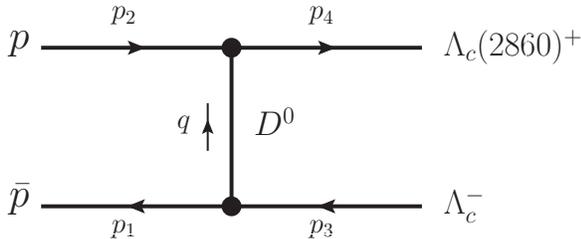}}
		\caption{The diagram describing the $\bar p p\to \Lambda_c^-\Lambda_c(2860)^+$ process.\label{fig:2to2}}
	\end{center}
\end{figure}

As discussed above, $\Lambda_c(2860)^+$ could be produced in the antiproton and proton collision by exchanging a $D^0$ meson, as shown in Fig. \ref{fig:2to2}. It should be noted that the $p\bar{p}$ annihilation ($s$ channel) is Okubo-Zweig-Iizuka (OZI) suppressed process. Thus, the contribution from the annihilation channel can be negligible compared with the contribution from the $t$ channel shown in Fig. \ref{fig:2to2}. In our calculation, only the $t$ channel is considered. 

\begin{figure}[htb]
	\begin{center}
		\scalebox{0.9}{\includegraphics[width=\columnwidth]{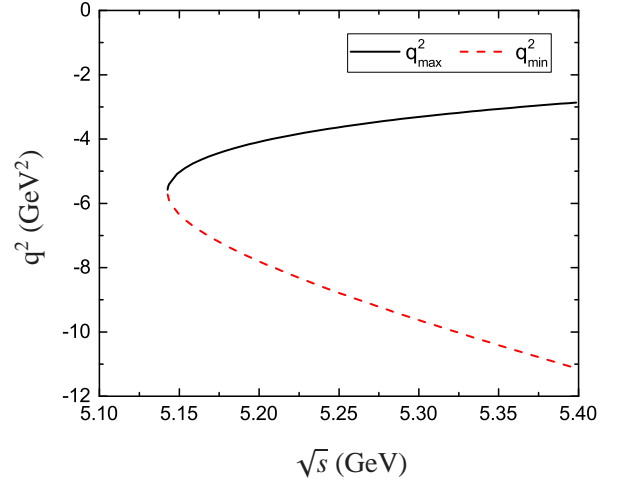}}
		\caption{The kinematically allowed region for the momentum of the transfer momentum in the processes $p\bar p \to \Lambda_c(2860)^+\Lambda_c^-$.
			\label{fig:qSuqre}}
	\end{center}
\end{figure}

Before evaluating the cross section of $\bar p p\to \Lambda_c^-\Lambda_c(2860)^+$, we display the kinematically allowed region of the square of the transfer momentum $q^2$ in Fig. \ref{fig:qSuqre}, which is the function of the center-of-mass (c.m.) energy $\sqrt{s}$. As shown in Fig. \ref{fig:qSuqre}, the maximum of $q^2$ is negative and less than the mass square of the exchanged $D^0$ meson in the energy range of our interest.

\subsection{The formalism}\label{sec21}

The effective Lagrangian approach is utilized to sudy the $\bar{p}p \to \bar{\Lambda}_c\Lambda_c(2860)^+$ process. As measured by LHCb \cite{Aaij:2017vbw}, we take the quantum number of $\Lambda_c(2860)^+$ to be $J^{P}={3}/{2}^+$. To describe the interaction of the nucleon with the charmed meson and the charmed baryon, we adopt the following effective Lagrangians \cite{He:2011jp,Dong:2009tg,Dong:2010xv,Xie:2015zga,Huang:2016ygf,Wu:2009md}
\begin{eqnarray}
	\mathcal{L}_{Dp\Lambda_c} &=& g_{D^0p\Lambda_c}\overline{\Lambda}_c i\gamma_5 D^0p + \rm{H.c.},\label{eq:lagrangianL} \\
	\mathcal{L}_{D^{*}p\Lambda_c} &=& g_{D^{*}p\Lambda_c}\overline{\Lambda}_c \gamma^\mu D_\mu^{*0}p + \rm{H.c.},\label{eq:lagrangianLs} \\
	\mathcal{L}_{DN\Sigma_c} &=& -g_{D^0N\Sigma_c}\overline{N} i\gamma_5 \bm{\tau}\cdot\bm{\Sigma_c}\overline{D} + \rm{H.c.},\label{eq:lagrangianS} \\
	\mathcal{L}_{D^{*}N\Sigma_c} &=& g_{D^{*}N\Sigma_c}\overline{N} \gamma^\mu \bm{\tau}\cdot\bm{\Sigma_c}\overline{D}_\mu^{*} + \rm{H.c.},\label{eq:lagrangianSs} \\
	\mathcal{L}_{DpR} &=& g_{D^0pR}\overline{R}^{\mu}\partial_\mu D^0p + \rm{H.c.},\label{eq:lagrangianR}
\end{eqnarray}
where $p/N$, $ \Lambda_c $, $ \Sigma_c $, $R$, $D^0$ and $D^{*0}$ denote the proton/nucleon, $ \Lambda_c(2286)^+ $, $ \Sigma_c(2455)^+ $, $\Lambda_c(2860)^+$, $D^0$ and $D^{*0}$ fields, respectively. In the following formulae, the abbreviations $g_{\Lambda_c}\equiv g_{Dp\Lambda_c}$, $ g_{\Lambda_c}^{\prime}\equiv g_{D^{*}p\Lambda_c} $, $g_{\Sigma_c}\equiv g_{DN\Sigma_c}$, $ g_{\Sigma_c}^{\prime}\equiv g_{D^{*}N\Sigma_c} $ and $g_R\equiv g_{DpR}$ are implemented. The coupling constants $ g_{\Lambda_c}=-13.98 $, $ g_{\Lambda_c}^{\prime}=-5.20 $, $ g_{\Sigma_c}=-2.69 $, $ g_{\Sigma_c}^{\prime}=3.0 $ are determined from the SU(4) invariant
Lagrangians in terms of $ g{\pi NN}=13.45 $ and $ g_{\rho NN}=6.0 $ \cite{Dong:2009tg,Dong:2010xv,Xie:2015zga}. The coupling constant $ g_R $ will be discussed later.
	
The propagators for the fermion with $J=1/2$, and $3/2$ are expressed as \cite{Machleidt:1987hj,Tsushima:1998jz,Huang:2005js,Wu:2009md}
\begin{eqnarray}
	G_\mathcal{F}^{n+(1/2)}(p) &=& P^{(n+(1/2))}(p)\frac{i2m_\mathcal{F}}{p^2-m_\mathcal{F}^2+im_\mathcal{F}\Gamma_\mathcal{F}}
\end{eqnarray}
with
\begin{eqnarray}
	P^{1/2}(p) &=& \frac{\slashed{p}+m_\mathcal{F}}{2m_\mathcal{F}},\\
	P^{3/2}(p) &=& \frac{\slashed{p}+m_\mathcal{F}}{2m_\mathcal{F}}Q_{\mu\nu}(p),\label{eq:proj5}\\
	Q_{\mu\nu}(p) &=& -g_{\mu\nu}+\frac{1}{3}\gamma_{\mu}\gamma_{\nu}+\frac{1}{3m_{\mathcal{F}}}(\gamma_{\mu}p_{\nu}-\gamma_{\nu}p_{\mu}) \nonumber\\
	&& +\frac{2}{3m^2_{\mathcal{F}}}p_{\mu}p_{\nu},
\end{eqnarray}
where $p$ and $m_\mathcal{F}$ are momentum and mass of the fermion, respectively. The propagators for the exchanged $ D^0 $ and $ D^{*0} $ are written as
\begin{eqnarray}
	G_D(q^2) &= \frac{i}{q^2-m_D^2} \\
	G_{D^{*0}}(q^2) &= \frac{i\left(-g^{\mu\nu}+\frac{q^\mu q^\nu}{q^2}\right)}{q^2-m_{D^{*0}}^2}
\end{eqnarray}

In the effective Lagrangian approach, the cross section of $p\bar{p}\to\bar{\Lambda}_c\Lambda_c(2860)^+$ is proportional to $g_R^2$ and the line shape depends on the c.m. energy. Here, a concrete $g_R$ value is adopted to execute the calculations. Generally, the coupling constant $g_{R}$ can be obtained by fitting the measured partial width of the $\Lambda_c(2860)^+(k) \to D^0(q)p(p)$ decay, where the partial decay width is
\begin{eqnarray}
	d\Gamma_i &=& \frac{m_Rm_N}{8\pi^2}|\mathcal{M}|^2\frac{|\vec{q}|}{m_R^2}d\Omega
\end{eqnarray}
with
\begin{eqnarray}
	E_q &=& \frac{m_R^2-m_N^2+m_D^2}{2m_R}, \\
	|\vec{q}| &=& \frac{\sqrt{\left[m_R^2-(m_D+m_N)^2\right]\left[m_R^2-(m_D-m_N)^2\right]}}{2m_R}.
\end{eqnarray}
Here, $E_q$ and $\vec{q}$ denote the energy and the three-momentum of the daughter $D^0$ meson, respectively.
$m_N$ and $m_D$ are the masses of proton and $D^0$ meson, respectively.
Furthermore, the concrete expression of the corresponding decay width is
\begin{eqnarray}
	\Gamma_i &=& \frac{g_R^2 m_N|\vec{q}|}{8\pi}\sum Tr[u(p)\bar{u}(p)q_{\mu}u_R^{\mu}(k)\bar{u}_R^{\nu}(k)q_{\nu}] \nonumber\\
	&=& \frac{g_R^2|\vec{q}|}{16\pi m_R} Tr[(\slashed{p}+m_N)\tilde{P}^{3/2}(k)q_{\mu}q_{\nu}],
\end{eqnarray}
where $P^{3/2}(k)$ is the projection operator for a fermion with $J=3/2$ as defined in Eq. (\ref{eq:proj5}). Until now, the branching ratio of  $\Lambda_c(2880)^+ \to D^0p$ is still not measured experimentally. We notice that the branching ratio of $\Lambda_c(2860)^+ \to D^0p$ has been theoretically predicted in several previous work. Thus it's feasibly to determine the coupling constant $g_{R}$ by the theoretical result, where the branching fraction $BR(\Lambda_c(2860)^+ \to D^0p)=48\%$ estimated in Ref. \cite{Chen:2017aqm} is adopted. Considering the above situation, in this work we extract $g_{R}=10.25$ GeV$^{-1}$ by taking a typical value $\Gamma(\Lambda_c(2860)^+ \to D^0p)=32.4$ MeV. In addition, the discussion for the contribution of the width uncertainty is necessary. By considering the experimental systematic and model uncertainties, one may obtain the uncertainty to be $ _{-18.6}^{+7.3} $ MeV on the total width. Here, we find that the positive one $ +7.3 $ MeV is about ten percent of the total width $ 67.6 $ MeV. To estimate the effect of the uncertainties of the coupling constant on the cross section, we suppose the total width has a margin of error of plus or minus 10 percentage points. The value of $ g_R $ is in the range of $9.73-10.76$ GeV$^{-1}$. These results are applied to the following calculations.

Additionally, since the hadrons are not pointlike particles, the monopole form factor \cite{He:2011jp,Xie:2015zga}
\begin{equation}
	\mathcal{F}_M(q^2,m^2)=\frac{\Lambda^2-m^2}{\Lambda^2-q^2}
\end{equation}
is introduced to phenomenologically describe the inner structure effect of the interaction vertices and compensates the off-shell effect for the $ t $ channel with the $ D^0 $ or $ D^{*0} $ meson exchange. Meanwhile, a form factor \cite{Shklyar:2005xg}
\begin{equation}
	\mathcal{F}_B(q^2,m^2)=\frac{\Lambda^4}{\Lambda^4+(q^2-m^2)^2}
\end{equation}
is also employed for the intermediate baryons. Here, the $ q $ and $ m $ denote the four-momentum and mass of the exchanged hadron, respectively. The cutoff $ \Lambda $ will be discussed below.

\subsection{The cross section of $p\bar{p} \to \Lambda_c^-\Lambda_c(2860)^+$}\label{sec22}

The transition amplitude for the process $p\bar{p} \to \Lambda_c^-\Lambda_c(2860)^+$ as shown in Fig. \ref{fig:2to2} can be expressed as
\begin{eqnarray}\label{eq:2to2}
	\mathcal{M} &=& \bar{u}_R(p_4)\mathcal{V}_R(q)u_{p}(p_2)\bar{v}_{\bar{p}}(p_1)\mathcal{V}(q)v_{\bar{\Lambda}_c}(p_3) \nonumber\\
	&&  \times G_D(q^2)\mathcal{F}_M^2(q^2,m_D^2),
\end{eqnarray}
where $ p_1 $, $ p_2 $, $ p_3 $, $ p_4 $ and $ q $ are the momenta of $ \bar{p} $, $ p $, $ \Lambda_c $, $ \Lambda_c(2860)^+ $ and the exchanged $ D^0 $ meson, respectively. $ \mathcal{V}_R $ or $ \mathcal{V} $ describes the Lorentz structure of the $ \Lambda_c(2860)^+pD^0 $ or $ \bar{\Lambda}_c\bar{p}D^0 $ interaction vetex including coupling constant. They can be derived by virtue of the Lagrangians in Eqs. (\ref{eq:lagrangianL}) and (\ref{eq:lagrangianR}). 

The unpolarized cross section is \cite{Zyla:2020zbs}
\begin{eqnarray}
	\frac{d\sigma}{dt} &=& \frac{m_Nm_Nm_{\Lambda_c}m_R}{16\pi s}
	\frac{1}{|\vec{p}_1|^2}\sum|\mathcal{M}|^2,
\end{eqnarray}
where
\begin{eqnarray}
	\sum|\mathcal{M}|^2 &=& |G_D(q^2)|^2\mathcal{F}_M^4(q^2,m_D^2) \nonumber\\
	&&  \times Tr\left[P^{3/2}(p_4)\mathcal{V}_R(q)\frac{\slashed{p}_2+m_N}{2m_N}\gamma^0\mathcal{V}_R(q)^\dag\gamma^0\right] \nonumber\\
	&&  \times Tr\left[\frac{\slashed{p}_1-m_N}{2m_N}\mathcal{V}\frac{\slashed{p}_3-m_{\Lambda_c}}{2m_{\Lambda_c}}
	\gamma^0\mathcal{V}^\dag\gamma^0 \right].
\end{eqnarray}

\begin{figure}[htb]
	\begin{center}
		\includegraphics[width=\columnwidth]{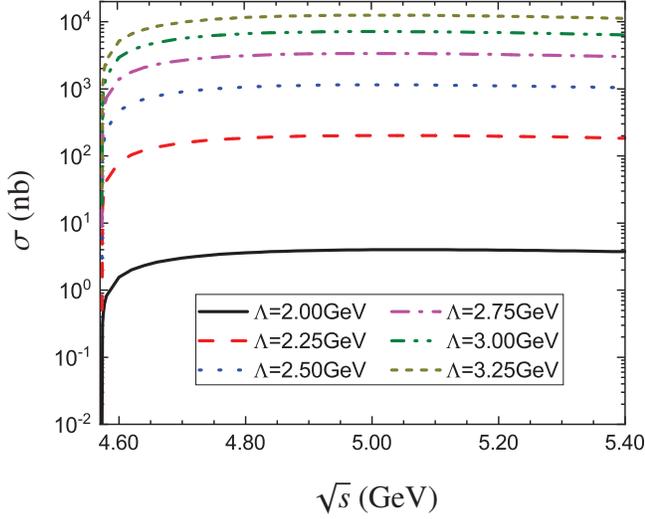}
		\caption{The obtained total cross section for $p\bar p \to \Lambda_c^-\Lambda_c^+$ with different cutoff values.
			\label{fig:TCS2toLL}}
	\end{center}
\end{figure}

Before studying the cross section for the $ \Lambda_c(2860)^+ $, it is necessary to calculate the total cross section for the reaction $p\bar{p}\to \Lambda_c^-\Lambda_c^+$. It is of great importance for the background analyses. The transition amplitude of $p\bar{p}\to \Lambda_c^-\Lambda_c^+$ can be obtained by replacing $\mathcal{V}_R(q)$ with $\mathcal{V}(q)$ in Eq. (\ref{eq:2to2}). In Fig. \ref{fig:TCS2toLL}, the total cross section of $p\bar{p}\to \Lambda_c^-\Lambda_c^+$ with different cutoff values is presented. The cutoff $ \Lambda $ in the form factor is a phenomenological parameter and we restrict the $ \Lambda $ value within a reasonable range from 2.00 GeV to 3.25 GeV.

\begin{figure}[htb]
	\begin{center}
		\includegraphics[width=\columnwidth]{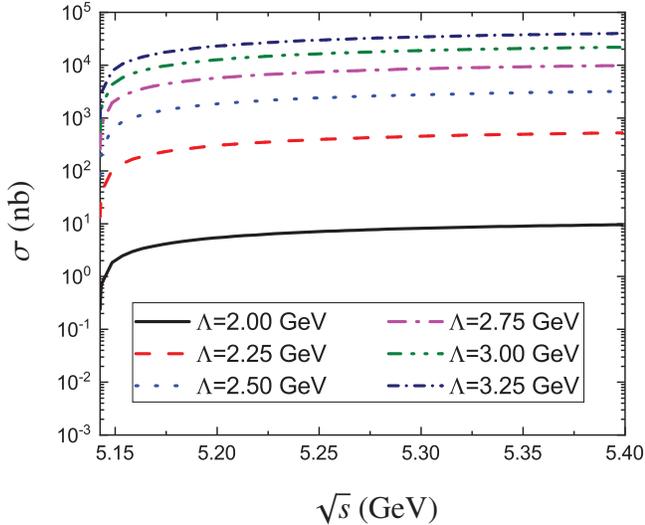}
		\caption{The obtained total cross section for $p\bar p \to \Lambda_c^-\Lambda_c(2860)^+$ with different cutoffs.
			\label{fig:TCS2toLR}}
	\end{center}
\end{figure}

With the similar consideration, the cross sections for the production of $ \Lambda_c(2860)^+ $ with different cutoffs are presented in Fig. \ref{fig:TCS2toLR}. Our results indicate that the cross section strongly depends on the values of $ \Lambda $. The cross section with $ \Lambda=2.00 $ is much smaller than that with $ \Lambda=3.25 $ by a fraction of $ \sim 10^4 $. In addition, the cross section of $p\bar{p}\to \Lambda_c^-\Lambda_c^+$ is smaller than that of $p\bar{p}\to \Lambda_c^-\Lambda_c^+(2860)$ if taking a same cutoff. This is mainly due to the reason that the exchanged $ \Lambda_c^+ $ is off-shell while the $ \Lambda_c^+(2860) $ can be on-shell.

In Ref. \cite{He:2011jp}, the production of $ \Lambda_c(2940)^+ $ via the $\bar{p}p$ collision was studied, where the cutoff was set to be 3~GeV. Considering the obvious similarity between these reactions, we adopt the same value to estimate the production rate of $ \Lambda_c(2860)^+ $ in the $ \bar{p}p $ reaction. In addition, we use other cutoff parameters $ \Lambda_{D^{*}}=\Lambda_{\Lambda_c}=\Lambda_{\Lambda_c^*}=\Lambda_{\Sigma_c}=\Lambda=3.0 $GeV for minimizing the free parameters.

\begin{figure}[htb]
	\begin{center}
		\scalebox{0.9}{\includegraphics[width=\columnwidth]{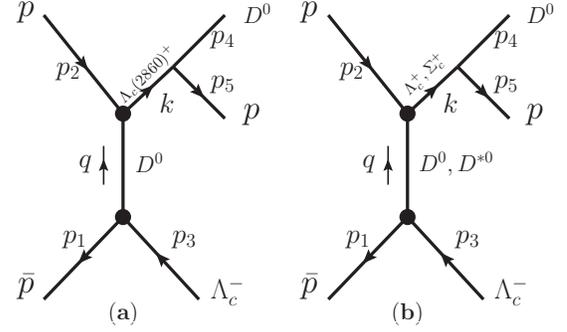}}
		\caption{The diagrams for $p\bar{p} \to D^0p\bar{\Lambda}_c$
			with different intermediate particles. (a) The signal channel with exchanged $ D^0 $ meson and intermediate $\Lambda_c(2860)^+$ contributions. (b) The background with exchanged $ D^0/D^{*0} $ mesons and intermediate $ \Lambda_c/\Sigma_c $ contributions.
			\label{fig:2to3}}
	\end{center}
\end{figure}

\section{The background analysis}\label{sec3}

The background analysis and invariant mass spectrum are also important for the study of $ \Lambda_c(2860)^+ $ production in $ \bar{p}p $ reaction. They may give more information for the corresponding reaction. In this section, we present Dalitz plot and $ pD^0 $ invariant mass spectrum for the reaction $p\bar p \to \Lambda_c^-\Lambda_c(2860)^+$ as shown in Fig. \ref{fig:2to3}, where the intermediate state $ \Lambda_c(2860)^+ $, $ \Lambda_c(2286)^+ $ and $ \Sigma_c(2455)^+ $ are involved. The processes $p\bar{p}\to\Lambda_c^-\Lambda_c^+\to\Lambda_c^-pD^0$ and $p\bar{p}\to\Lambda_c^-\Sigma_c^+\to\Lambda_c^-pD^0$ with both $ D^0 $ and $ D^{*0} $ exchanges are as the main background contributions.

The transition amplitude of $p\bar{p}\to\Lambda_c^-\Lambda_c(2860)^+\to\Lambda_c^-pD^0$ is written as
\begin{eqnarray}
	\mathcal{M}_{a} &=&  \bar{u}_p(p_4)\mathcal{V}_R(p_5)G_R^{3/2}(k)\mathcal{V}_R(q)u_{p}(p_2)G_D(q^2) \nonumber\\
	&& \times  \bar{v}_{\bar{\Lambda}_c}(p_3)\mathcal{V}_{\Lambda_c}(q)v_{\bar{p}}(p_1)\mathcal{F}_M^2(q^2,M_D^2) \nonumber\\
	&& \times \mathcal{F}_B(k^2,M_R^2), \label{eq:2to3pD}
\end{eqnarray}
The involved momenta are defined in Fig. \ref{fig:2to3}. One can easily obtain the amplitudes of the other four processes, as shown in Fig. \ref{fig:2to3} (b), by replacing the relevant masses, form factors, propagators and vertices which can be derived from Eqs.\ref{eq:lagrangianL}-\ref{eq:lagrangianR}. Here, the sum of the four amplitudes is expressed as $ \mathcal{M}_{b} $.

With the above amplitudes, the square of the total invariant transition amplitude reads as
\begin{eqnarray}
	|\mathcal{M}|^2 &=& \sum|\mathcal{M}_{a}+\mathcal{M}_{b}|^2.
\end{eqnarray}
The corresponding total cross section of the process $p\bar{p}\to \Lambda_c^-pD^0$ is
\begin{eqnarray}
	d\sigma &=& \frac{m_N^2}{|p_1\cdot p_2|}\frac{|\mathcal{M}|^2}{4}(2\pi)^4d\Phi_3(p_1+p_2;p_3,p_4,p_5)
\end{eqnarray}
with the definition of $n$-body phase space \cite{Zyla:2020zbs}
\begin{equation}
	d\Phi_n(P;k_1,...,k_n)=\delta^4(P-\sum\limits_{i=1}^nk_i)\prod\limits_{i=1}^3\frac{d^3k_i}{(2\pi)^32E_i}.
\end{equation}

\begin{figure}[htb]
	\begin{center}
		\scalebox{0.9}{\includegraphics[width=\columnwidth]{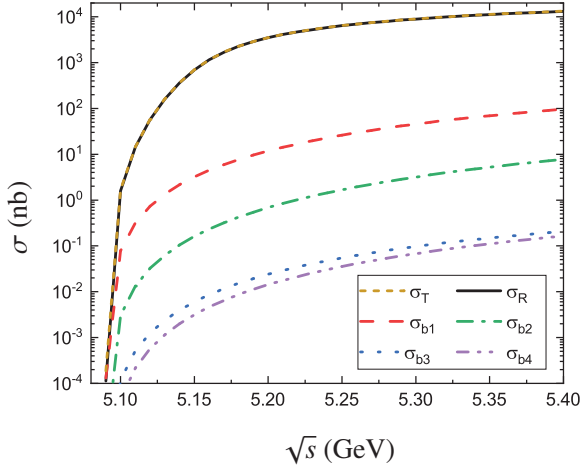}}
		\caption{The obtained total cross section for $p\bar p \to \Lambda_c^-pD^0$ with $ g_R=10.25 $ GeV$ ^{-1} $.\label{fig:TCS2toLpD}}
	\end{center}
\end{figure}

\begin{figure}[htb]
	\begin{center}
		\scalebox{0.45}{\includegraphics[width=\columnwidth]{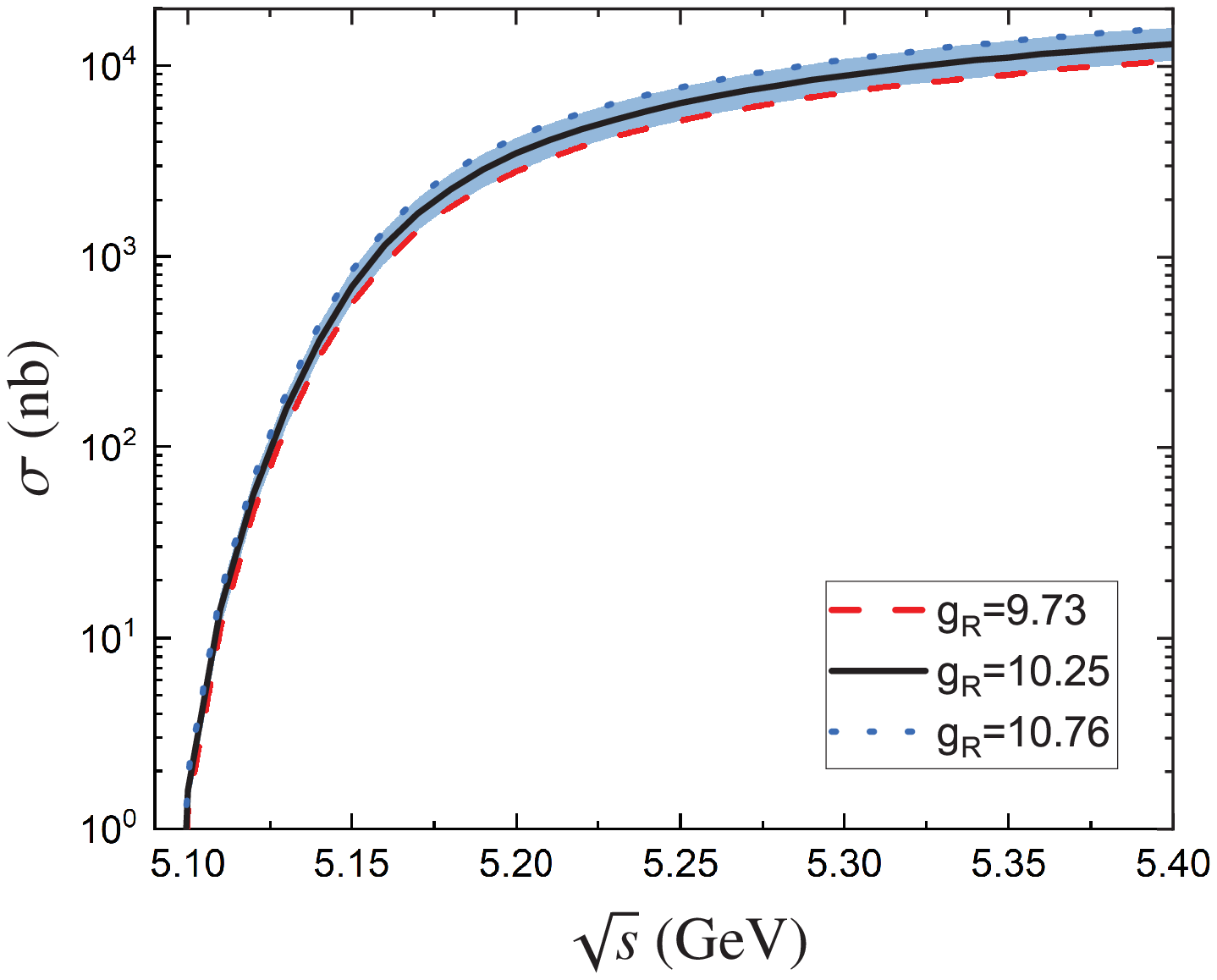}}
		\scalebox{0.45}{\includegraphics[width=\columnwidth]{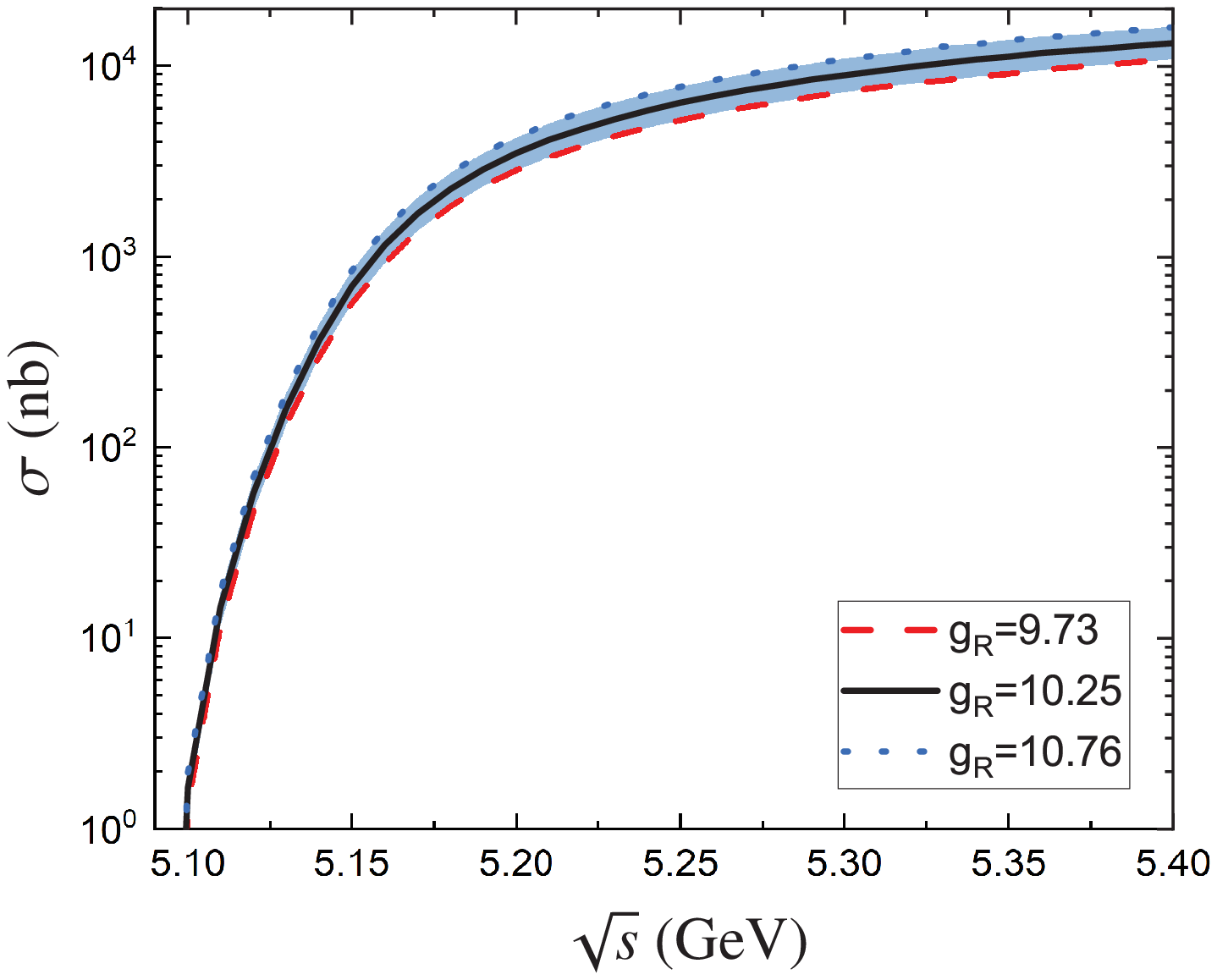}}
		\caption{The obtained total cross section for $p\bar p \to \Lambda_c^-pD^0$ with different $ g_R $. The left and right figures correspond to $ \sigma_R $ and $\sigma_{T}$, respectively.\label{fig:LpD}}
	\end{center}
\end{figure}

In Fig. \ref{fig:TCS2toLpD}, the total cross section is given, which is dependent on $ \sqrt{s} $. Here, $\sigma_R$ and $\sigma_{T}$ correspond to the signal and total cross section, respectively. $\sigma_{b1}$ and $\sigma_{b2}$ correspond to that of  $p\bar{p}\to\Lambda_c^-\Lambda_c^+\to\Lambda_c^-pD^0$ with $ D^0 $ and $ D^{*0} $ exchanges, respectively. $\sigma_{b3}$ and $\sigma_{b4}$ correspond to that of $p\bar{p}\to\Lambda_c^-\Sigma_c^+\to\Lambda_c^-pD^0$ with $ D^0 $ and $ D^{*0} $ exchanges, respectively. As shown in Fig. \ref{fig:TCS2toLpD}, $\sigma_{b1}+\sigma_{b2}$ is much larger than $\sigma_{b3}+\sigma_{b4}$, which indicates that the reaction $p\bar p \to \Lambda_c^-pD^0$ via the intermediate $ \Lambda_c^+ $ should be the main background. One can also find that $ \sigma_{b1} $ reaches up to about 100 nb at $ \sqrt{s}=5.4$ GeV, which is about one order of magnitude larger than $ \sigma_{b2} $. Thus the main contribution to the background comes from the raction $p\bar{p}\to\Lambda_c^-\Lambda_c^+\to\Lambda_c^-pD^0$ with the $ D^{0} $ exchange. The cross section for the production of $ \Lambda_c(2860)^+ $ increases rapidly near the threshold and then reaches up to about 13 $ \mu $b at $ \sqrt{s}=5.4 $ GeV. The results indicate that the signal can be easily distinguished from the background. In addition, if the total width of $ \Lambda_{c}(2860) $ has an uncertainty of 10\%, the coupling constant $ g_R $ is the range of $9.73-10.76$ GeV$ ^{-1} $. The corresponding cross sections with these different $ g_R $ values are shown in Fig. \ref{fig:LpD}.

\begin{figure}[htbp]
	\begin{center}
		\scalebox{0.45}{\includegraphics[width=\columnwidth]{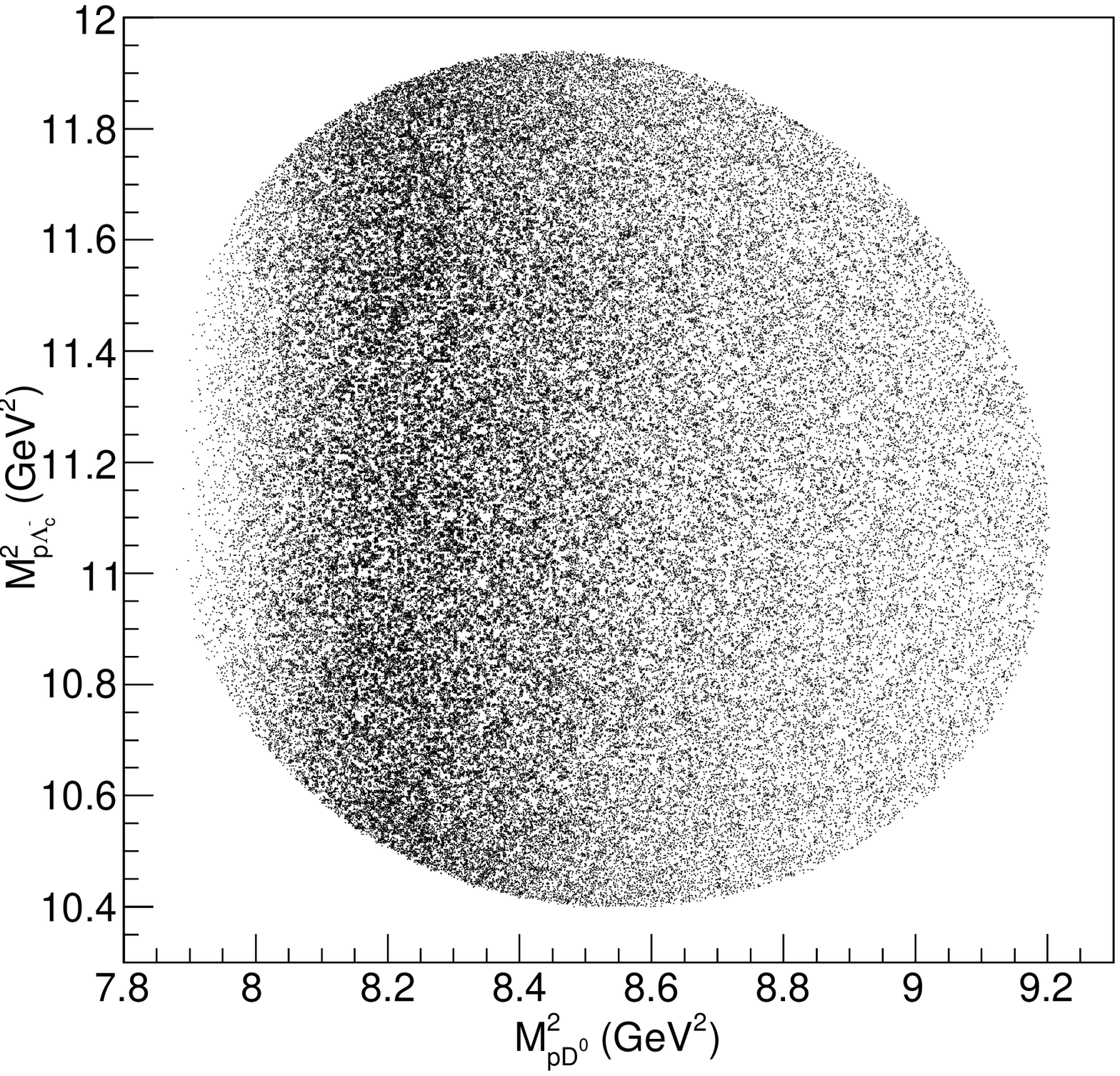}}
		\scalebox{0.45}{\includegraphics[width=\columnwidth]{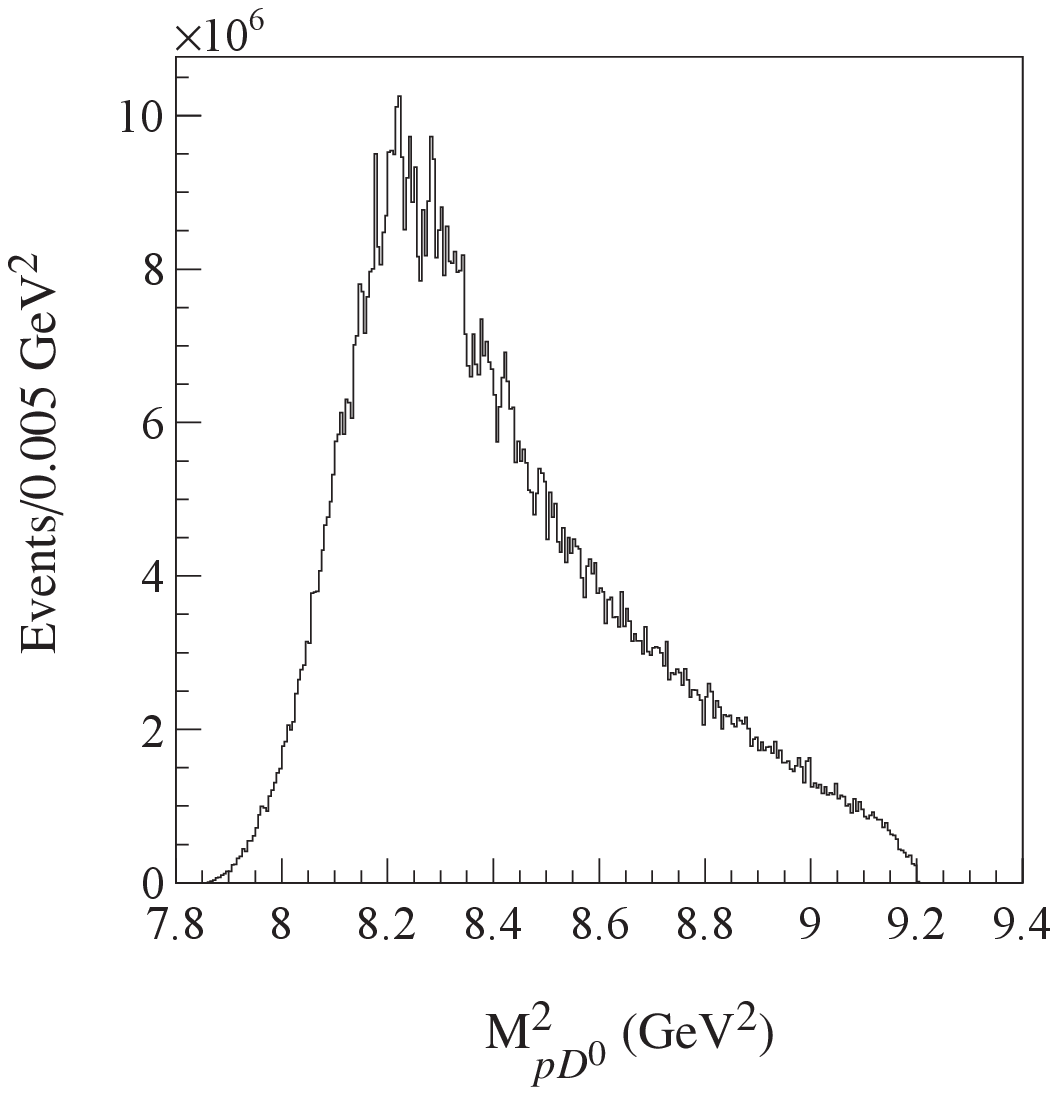}}
		\caption{The Dalitz plot(left) and $ pD^0 $ invariant mass spectrum distribution(right) for $p\bar p \to \Lambda_c^-pD^0$ at $ \sqrt{s}=5.32 $GeV.	\label{fig:dalitzpD}}
	\end{center}
\end{figure}

With the help of Mathemaica and FOWL codes, we present the Dalitz plot for the $p\bar p \to \Lambda_c^-pD^0$ process and the $ pD^0 $ invariant mass spectrum at $ \sqrt{s}=5.32 $GeV in Figs. \ref{fig:dalitzpD}.
	
\section{Summary}\label{sec4}

In this work, we investigate the production of the charmed baryon $\Lambda_c(2860)$ via antiproton-proton reaction, which is different from the $\Lambda_c(2860)$ production observed in the $\Lambda^0_b$ decay \cite{Aaij:2017vbw}. The present study can supply valuable information for the experimental search for $\Lambda_c(2860)$ in the future experiments at $\overline{\mbox{P}}$ANDA \cite{Lutz:2009ff}.

It should be noted that the initial state interaction (ISI) and final state interaction (FSI) may play an important role on the nucleon-nucleon entrance channel \cite{Hanhart:1998rn,Baru:2002rs,Dong:2014ksa,Haidenbauer:2009ad}. However, the ISI and FSI effects are thought to be described by the nonperturbative QCD and should be rather complicated. We notice that the authors in Refs. \cite{Hanhart:1998rn,Dong:2014ksa} studied the ISI effects on nucleon collisions. Their results indicate that the ISI leads to a suppression on the cross section, which may change the cross section by a factor no more than $ 10\%-15\% $. Furthermore, with the frame of the J$ \ddot{u} $lich meson-baryon model \cite{Haidenbauer:2009ad}, the FSI effect on the cross section was implemented. However, the aim of the present work is to carry out the discovery potential of $\Lambda_c(2860)$ produced at $\overline{\mbox{P}}$ANDA. The ISI and FSI effects are rather complicated and go beyond the scope of this work. Therefore, as suggested by previous research \cite{Baru:2002rs,He:2011jp}, an reasonable factor is introduced to reflect the ISI effect, which makes the cross section of $p\bar p \to \Lambda_c^-\Lambda_c(2860)^+$ suppressed by 1 order of magnitude (this factor is considered for the above calculations). With the above consideration, one can roughly estimate the events of $\Lambda_c(2860)$ produced at $\overline{\mbox{P}}$ANDA. Considering the designed luminosity of $\overline{\mbox{P}}$ANDA ($2\times10^{32}~\mbox{cm}^{-2}\mbox{s}^{-1}$), one may expect that there are about $ 10^8 $ $\Lambda_c(2860)$ events accumulated per day by reconstructing the final $ pD^0 $. The Dalitz plot and $ pD^0 $ invariant mass spectrum analyses are also performed. We find that the signal can be easily distinguished from the background.

\begin{figure}[htbp]
	\begin{center}
		\scalebox{0.9}{\includegraphics[width=\columnwidth]{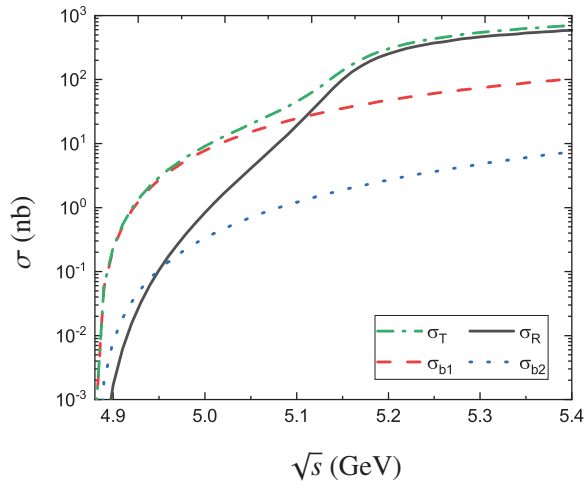}}
		\caption{The obtained total cross section for $p\bar p \to \Lambda_c^-\Sigma_c^{++}\pi^-$ with $ (g_R, g_R^\prime)=(10.25, 1.01) $.\label{fig:TCS2toLSpi}}
	\end{center}
\end{figure}

\begin{figure}[htb]
	\begin{center}
		\scalebox{0.45}{\includegraphics[width=\columnwidth]{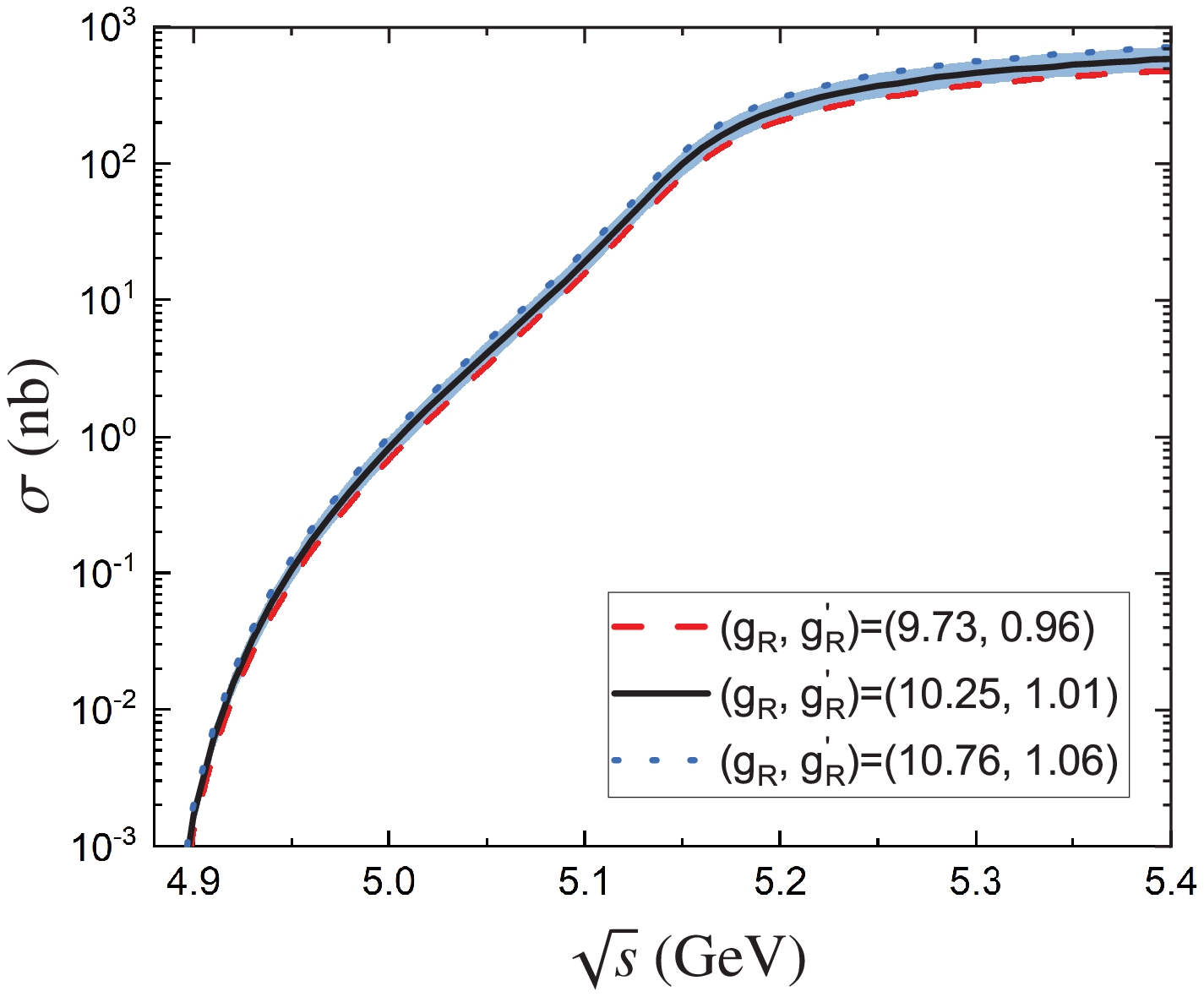}}
		\scalebox{0.45}{\includegraphics[width=\columnwidth]{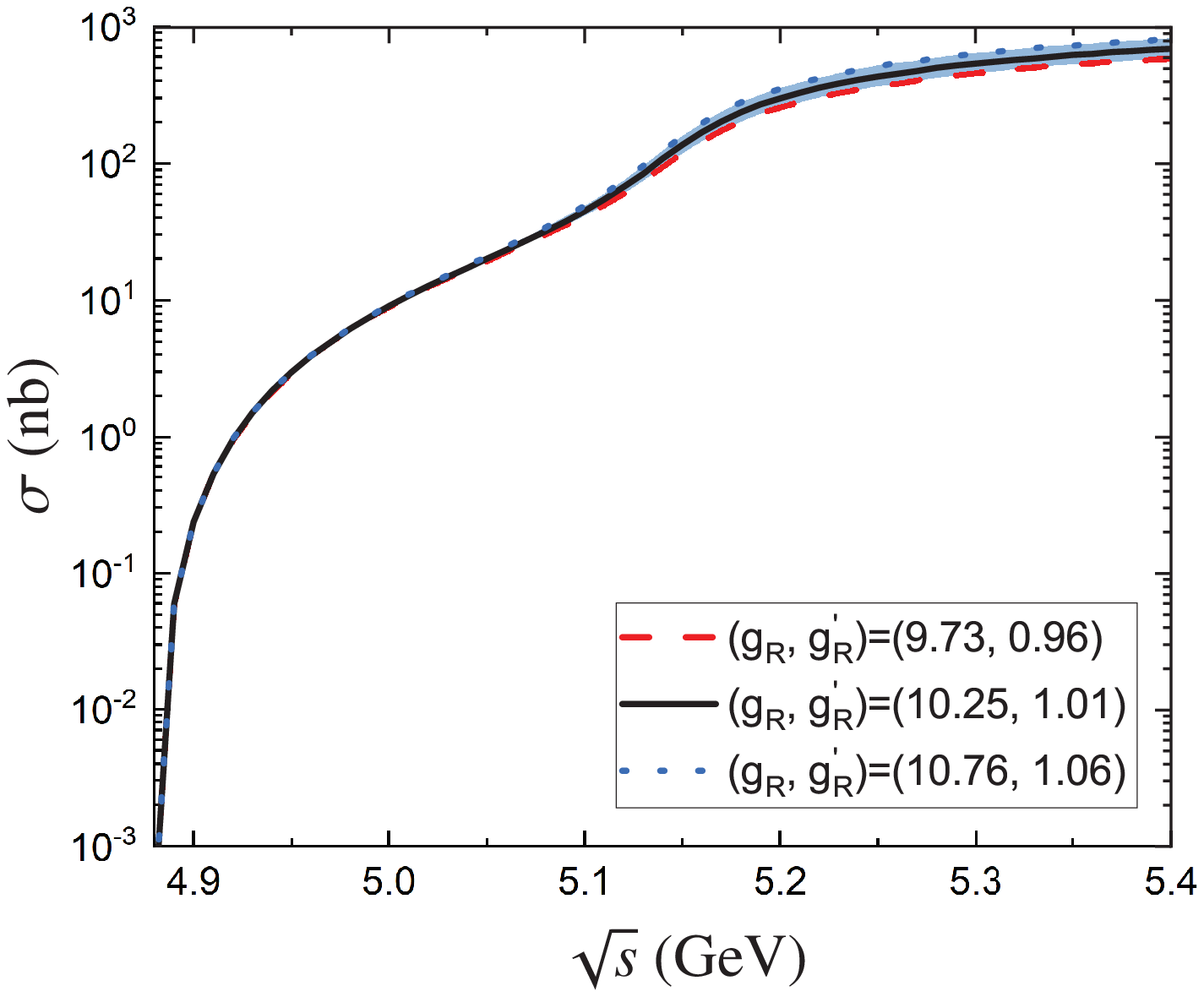}}
		\caption{The obtained total cross section for $p\bar p \to \Lambda_c^-\Sigma_c^{++}\pi^-$ with different $ g_R $ and $ g_R^\prime $. The left and right figures correspond to $ \sigma_R $ and $\sigma_{T}$, respectively.\label{fig:LSpi}}
	\end{center}
\end{figure}

\begin{figure}[htb]
	\begin{center}
		\scalebox{0.9}{\includegraphics[width=\columnwidth]{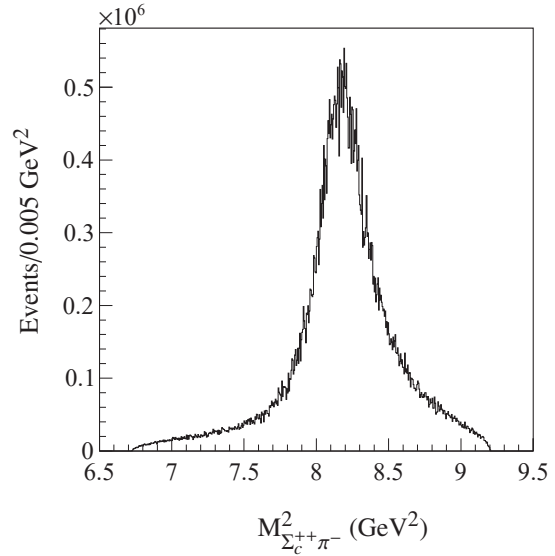}}
		\caption{The $ \Sigma_c^{++}\pi^- $ invariant mass spectrum distribution for $p\bar p \to \Lambda_c^-\Sigma_c^{++}\pi^-$ at $ \sqrt{s}=5.32 $GeV. \label{fig:specSpi}}
	\end{center}
\end{figure}

Inspired by the theoretical predictions \cite{Chen:2017aqm,Guo:2019ytq,Yao:2018jmc}, we also study the $\Lambda_c(2860)$ production in $p\bar p \to \Lambda_c^-\Sigma_c^{++}\pi^-$. Here, $p\bar p \to \Lambda_c^-\Lambda_c(2860)^+ \to \Lambda_c^-\Sigma_c^{++}\pi^-$ and $p\bar p \to \Lambda_c^-\Lambda_c^+ \to \Lambda_c^-\Sigma_c^{++}\pi^-$ correspond to the signal and background channel, respectively. For the background channel, the contributions are both from $ D^0 $ and $ D^{*0} $ exchanges, which are labeled as $ \sigma_{b1} $ and $ \sigma_{b2} $ in Fig. \ref{fig:TCS2toLSpi}. The cross section with $ BR(\Lambda_c(2860)^+ \to \Sigma_c^{++}\pi^-)\sim 3.0\% $ is presented in Fig. \ref{fig:TCS2toLSpi}, where the coupling constant $ g_{\Lambda_c\Sigma_c\pi}=9.32 $ is adopted \cite{Xie:2015zga}. As discussed above, the uncertainties of the coupling constants $ g_R $ and $ g_R^\prime\equiv g_{R\Sigma_c\pi} $ are also considered, with $ g_{R}^\prime $ varying from 0.96 to 1.06 GeV$ ^{-1} $. The corresponding results are shown in Fig. \ref{fig:LSpi}. The invariant mass spectrum of $ \Sigma_c^{++}\pi^- $ is also simulated and presented in Fig. \ref{fig:specSpi}. As shown in Figs. \ref{fig:TCS2toLSpi} and Fig. \ref{fig:specSpi}, the signal is several times larger than the background and can be distinguished clearly. Thus, the channel  $p\bar p \to \Lambda_c^-\Sigma_c^{++}\pi^-$ is also a suitable channel to study $\Lambda_c(2860)$. Due to the small branching fraction, we do not consider the contribution from $ \Sigma_c(2520)\pi $ channel.

In addition, as discussed in Ref. \cite{Yao:2018jmc}, the ratio $\mathcal{R}=\frac{\Gamma[\Sigma_c(2520)\pi]}{\Gamma[\Sigma_c(2455)\pi]}$ for the nearby state $\Lambda_c(2880)$ may be strongly affected by $\Lambda_c(2860)$. Thus, it is an interesting topic to study the $p\bar p \to \Lambda_c^-\Sigma_c(2455)^{++}\pi^-$ and $p\bar p \to \Lambda_c^-\Sigma_c(2520)^{++}\pi^-$ reactions within the contributions from both $\Lambda_c(2860)$ and $\Lambda_c(2880)$, which can be accessible at future experiment like $\overline{\mbox{P}}$ANDA.

\section*{Acknowledgments}

This project is supported by the National Natural Science Foundation of China (Grant Nos. 11747160), the Natural Science Foundation of Fujian Province (Grant No. 2018J05007) and the Natural Science Foundation of Jimei University (Grant No. ZQ2017007). XL is supported by the China National Funds for Distinguished Young Scientists under Grant No. 11825503, National Key Research and Development Program of China under Contract No. 2020YFA0406400, the 111 Project under Grant No. B20063, and the National Natural Science Foundation of China under Grant No. 12047501.

\vfill

\end{document}